\documentclass[9pt,twocolumn,twoside]{pnas-new}
\templatetype{pnasresearcharticle} 
\title{Ultrafast stratified diffusion of water inside carbon nanotubes. Direct experimental evidence with 2D \(D-T_2\) NMR spectroscopy}

\author[1]{J. Hassan}
\author[1]{G. Diamantopoulos} 
\author[2]{L. Gkoura}
\author[2]{M. Karayanni}
\author[3]{S. Alhassan}
\author[3]{S.V. Kumar}
\author[3]{M.S. Katsiotis}
\author[3]{T. Karagiannis}
\author[2]{M. Fardis}
\author[2]{N. Panopoulos}
\author[4]{H.J. Kim}
\author[5]{M. Katsioti}
\author[2]{G. Papavassiliou}

\affil[1]{Department of Physics, Khalifa University of Science and Technology, 127788, Abu Dhabi, UAE}
\affil[2]{Institute of Nanoscience and Nanotechnology, NCSR Demokritos, 15310 Aghia Paraskevi, Attiki, Greece}
\affil[3]{Department of Chemical Engineering, Khalifa University of Science and Technology, 2533, Abu Dhabi, UAE}
\affil[4]{Nano-Bio Electron Microscopy Research Group, Korea Basic Science Institute, 169-148, Daejeon 305-806, Republic of Korea}
\affil[5]{School of Chemical Engineering, National Technical University of Athens, 9 Iroon Polytechniou Street, 15780 Zografou, Athens, Greece}
\leadauthor{J. Hassan} 

\usepackage{epstopdf}
\begin{abstract}
Water, when confined at the nanoscale acquires extraordinary transport properties. And yet there is no direct experimental evidence of these properties at nanoscale resolution. Here, by using 2D NMR diffusion-relaxation (\(D-T_2\)) and spin-lattice - spin-spin relaxation (\(T_1-T_2\)) spectroscopy, we succeeded to resolve at the nanoscale water diffusion in single and double-walled carbon nanotubes (SWCNT/DWCNT). In SWCNTs, spectra display the characteristic shape of uniform water diffusion restricted in one dimension. Remarkably, in DWCNTs water is shown to split into two axial components with the inner one acquiring unusual flow properties: high fragility, ultrafast self-diffusion coefficient, and ``rigid'' molecular environment, revealing a stratified cooperative motion mechanism to underlie fast diffusion in water saturated CNTs.  
\end{abstract}

\dates{This manuscript was compiled on \today}
\begin{document}

\verticaladjustment{-2pt}

\maketitle
\thispagestyle{firststyle}
\ifthenelse{\boolean{shortarticle}}{\ifthenelse{\boolean{singlecolumn}}{\abscontentformatted}{\abscontent}}{}

The behavior of water in the vicinity of nanoscaled surfaces and porous structures diverges significantly from that of bulk water. For example, water in contact with graphene or carbon nanotubes (CNTs) shows astonishingly high transport properties \cite{Whitby2007}, with recent molecular dynamics simulations (MDS) unveiling new exotic physics as the origin of this intriguing behavior. Specifically, the ultrafast diffusion of water nanodroplets on graphene layers was recently shown to be driven by propagating ripples in a motion resembling surfing on sea waves  \cite{Ma2016a}. Similarly, water molecules in CNTs were predicted to be coupled with the longitudinal phonon modes of the nanotubes, leading to enhanced diffusion of the confined water by more than 300\%  \cite{Ma2015b}.  In the same consensus, membranes consisting of macroscopically aligned CNTs with diameters as small as $2$ nm are expected to exhibit  orders of magnitude higher water flow compared to predictions deriving from continuum hydrodynamics and Knudsen diffusion models \cite{Whitby2007}. 

In sub-nanometer narrow CNT channels, MDS have shown that water forms an ordered one-dimensional chain, which spontaneously fills the hydrophobic CNT channels \cite{hummer2001water}. The estimated flow rates are comparable to those observed in biological pores such as aquaporin, gramicidin, and bacteriorhodosin \cite{kong2001dynamic,pomes2002molecular,lanyi2000molecular}. Other MDS have shown that by increasing the CNT diameter, water inside the CNTs forms ice nanotubes, comprised of a rolled sheet of cubic ice \cite{koga2001formation}, whilst by further increasing the CNT channel size, neutron scattering experiments \cite{kolesnikov2004anomalously} and MDS \cite{bordin2013relation,joseph2008carbon,striolo2005water} have shown that water molecules are organized in a stratified nanotubular structure with a "chain"-like water component at the center of the CNTs. This intriguing molecular arrangement is central in explaining the fast water diffusion in CNTs \cite{kolesnikov2004anomalously,holt2006fast}, which is of critical importance in many applications. Today, it is generally considered that fast dynamics in hydrophobic nanochannels such as CNTs takes place either in the form of a single file \cite{kalra2003osmotic}, where water molecules can not surpass each other due to space limitation, or in a ballistic way, i.e. very fast clusters of hydrogen bonded water molecules diffuse in a highly coordinated way \cite{hummer2001water,striolo2006mechanism}. However, most of the facts regarding the nature of water diffusion in nanopores at molecular scale have been almost solely acquired by MDS studies; until now  experimental work is very scarce.

\begin{figure*}[t]
\centering
\includegraphics[width=\textwidth]{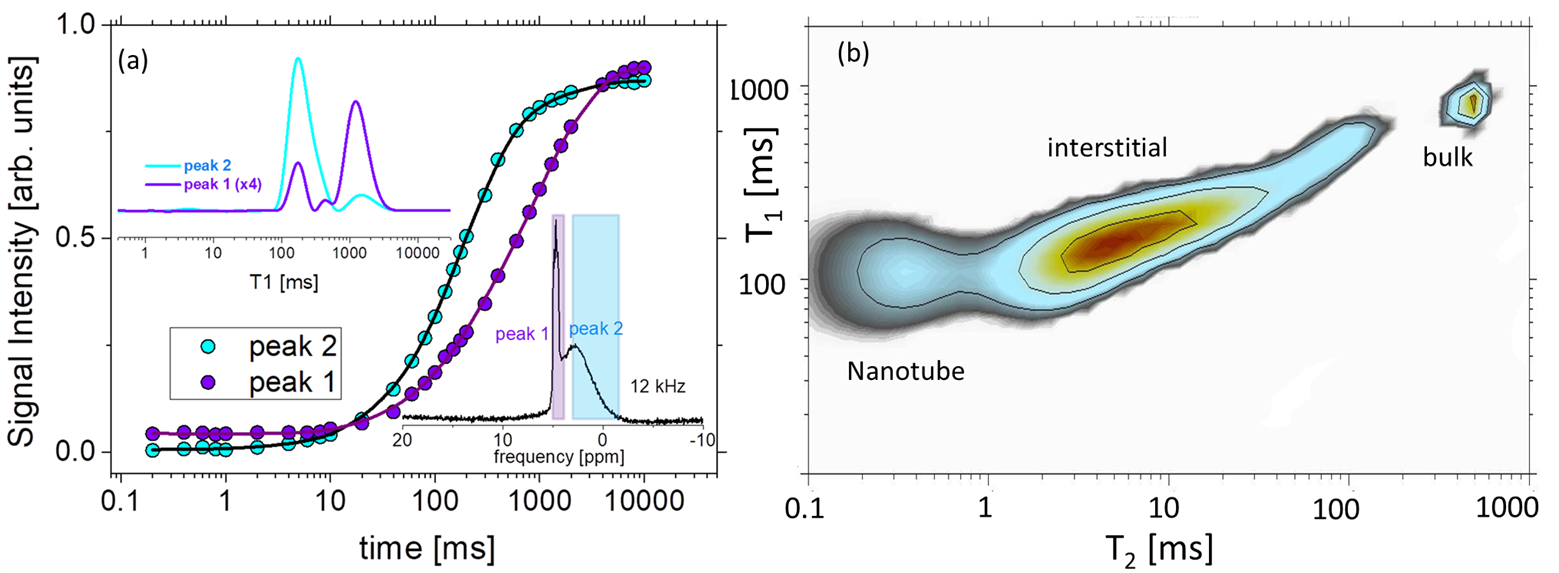}
\caption{(a)$^1$H MAS NMR $T_1$ analysis of water  in DWCNTs, at room temperature. The right inset shows the relevant $^1$H MAS NMR spectrum at spinning-frequency of $12$kHz. Peak $1$ corresponds to bulk water and the broad peak $2$ to nanotube/interstitial water. The main panel shows $^1$H MAS NMR relaxation data from the shaded areas of the spectrum (saturation recovery curves). The solid lines are theoretical fits by using an 1D inversion algorithm. The left inset shows the relevant $T_1$ distributions of bulk and confined water, respectively, obtained by the inversion. (b) Contour plot of the static (no MAS)$^1$H NMR $T_1-T_2$ spectrum of water in DWCNTs, at room temperature. Based on the $T_2/T_1$ ratios, bulk, interstitial, and nanotube water can be resolved.}
\label{fig:fig1}
\end{figure*}

Here, on the basis of  2D  $^1$H NMR diffusion-relaxation ($D-T_2$) and  relaxation ($T_1-T_2$) spectroscopy, we succeeded in resolving experimentally at the nanoscale the distribution of the self-diffusion coefficient $D$ of water inside SWCNTs and DWCNTs. Our experiments show that in SWCNTs,  the  (\(D-T_2\)) NMR spectra exhibit the characteristic shape of a uniform water diffusion in randomly oriented one dimensional nanochannels, with $D$ values similar to those of free water. Most important, diffusion in DWCNTs becomes non-uniform; a second axial water component is observed, with $D$ values four times that of bulk water at $T=285$K. To the best of our knowledge, this is the first direct experimental evidence of stratified water diffusion in CNTs.  It is furthermore noticed that the time window of the NMR diffusion measurements is $2-3$ orders of magnitude longer than in MDS. This allows assessment even of ultra-slow motion, such as of macromolecules through biological membranes, which makes the presented methodology applicable in important biological processes.
\section*{Results and Discussion}
\indent
Fig. \ref{fig:fig1}a, shows the $^{1}$H NMR $T_1$ distribution analysis of water in DWCNTs. The right inset shows the relevant $^1$H magic angle spinning (MAS) NMR spectrum, obtained at spinning rate  of $12$ kHz. Two peaks are observed, in agreement with a previous work \cite{Sekhaneh2006}; a broad one at $2.73$ ppm, which is attributed to water inside the CNTs (nanotubular water) and possibly to water in the space between the CNT bundles (interstitial water) and a relatively narrow peak at $4.9$ ppm, very close to the frequency of  bulk water, $4.8$ ppm. The frequency shift of the former signal from the frequency of the bulk water may be assigned either to less number of hydrogen bonds per water molecules inside the CNTs, or to shielding from ring currents induced on the CNT walls \cite{Sekhaneh2006}. The main panel shows the experimental relaxation data acquired separately on each of the two peaks, by using a saturation recovery technique \cite{Katsiotis2015a}. The $T_1$ analysis was then performed by implementing an 1D inversion algorithm on the  relaxation data (left inset of Fig.  \ref{fig:fig1}a) \cite{mitchell2012numerical}. Details on the inversion can be found in the Supporting Information. It is observed that bulk water acquires $T_1$ distribution with peak at $\approx1.2$ s (violet line corresponding to the shaded area around $4.9$ ppm), while the nanotube/interstitial water holds $T_1$ distribution with a peak at $\approx 0.15$ s (cyan line corresponding to the shaded area around $2.73$ ppm). Both values are sufficiently shorter than that of distilled bulk water $T_1 \approx 2$ s, due to the unavoidable presence of paramagnetic impurities on the CNT walls. The weak violet $T_1$ peak at $\approx0.15$ s belongs to the tail of the nanotubular water NMR signal, which overlaps with the NMR signal of the bulk water. In addition, a second weak $T_1$ peak at $\approx1.2$ s is observed in the cyan line, which shows that a small component of the confined water acquires bulk water dynamics. 

To further examine different  water groups in CNTs,  $^{1}$H NMR $T_1-T_2$ correlation spectroscopy was implemented under static conditions (no MAS), as shown in Fig. \ref{fig:fig1}b. 2D $T_1-T_2$ provides information about the ``liquidity'' of the local molecular environment: in general, molecules in unconstraint liquid environment are characterized by \(T_{2}\approx T_{1}\), whereas molecules in a rigid (solid) environment show \(T_2<<T_1\). In this context, bulk water with \(T_{2}\approx T_{1}\), semi-free interstitial water with \(T_2<T_1\), and solid-like nanotube water with \(T_2<<T_1\) can be distinguished in Fig.  \ref{fig:fig1}b.  It is stressed that according to MDS the interaction of water with the CNT walls depends on the wall curvature; in the outer space, water molecules are shown to interact moderately with the external CNT surfaces, whereas inside the CNTs water molecules do not interact with the CNT walls \cite{Falk2010,Lei2016}.  In this context, the very short  \(T_2/T_1<<1\) ratio of water inside the  CNTs indicates that nanotubular water molecules are in a rigid molecular environment. This evidence corroborates with the predictions from MDS, which show that water molecules in CNTs are organized in hydrogen bonded nanotubular clusters \cite{koga2001formation,bordin2013relation}.   

\begin{figure}[!ht]
\centering
\includegraphics[width=\linewidth]{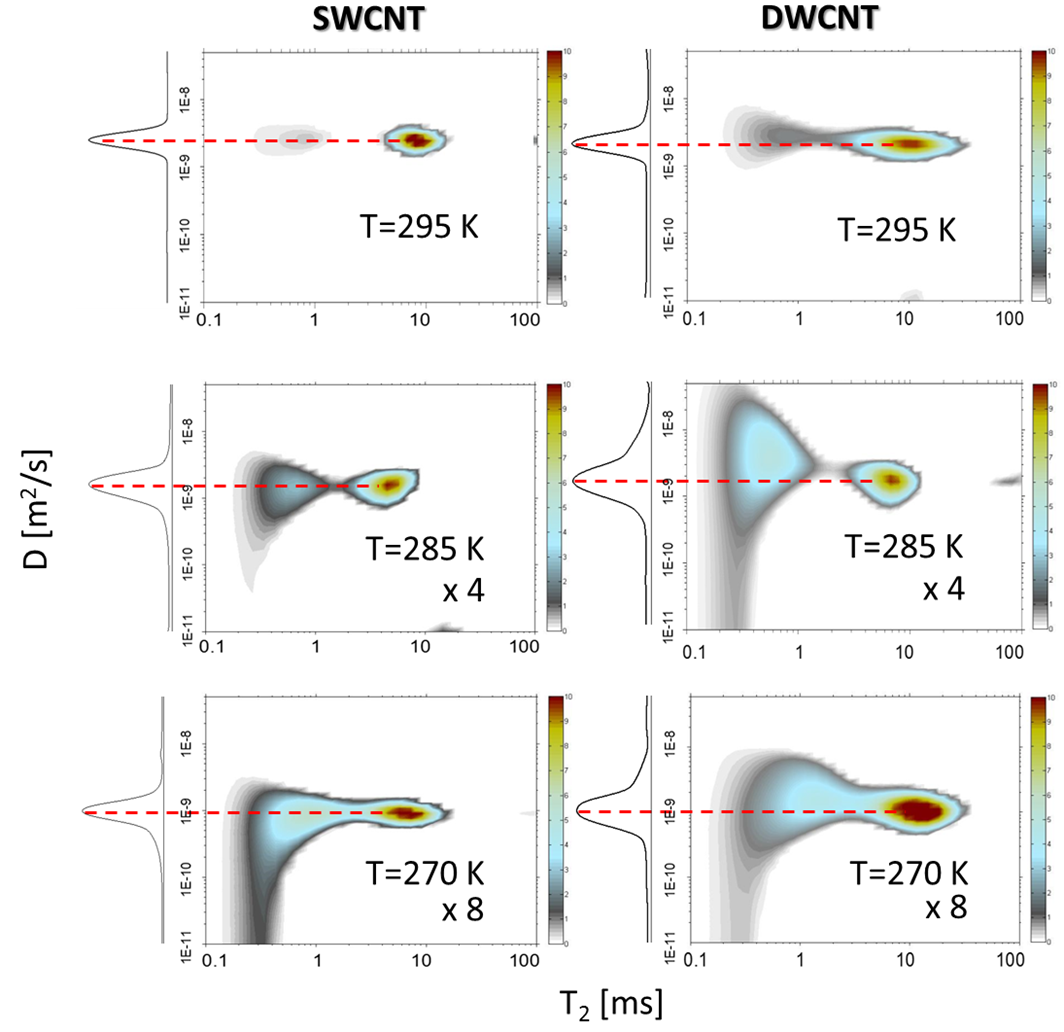}
\caption{Contour plots of $^{1}$H NMR \(D-T_2\) spectra of water in SWCNT 
(left panel) and DWCNT (right panel) at selected temperatures. Two main 
components are seen at each temperature, corresponding to nanotube water 
confined in the CNTs (low $T_2$ values) and interlayer and bulk water 
external to the CNTs (high $T_2$ values). Plots at the left side of the 
contour plots are the relevant $D$-projections.}
\label{fig:fig2}
\end{figure}

Static 2D (\(D-T_2\))  NMR experiments were performed in the stray field of a superconductive magnet with a constant magnetic field gradient \(G=34.7\) T/m. The conventional 1D NMR method for measuring water diffusion in a constant magnetic field gradient is to monitor the $^{1}$H NMR spin-echo decay and then fitting the obtained decay curve to relation $M=M_0\:exp{\left( { - \frac{{2\tau }}{{T_2 }} - \frac{2}{3}D\gamma ^2 G^2 \tau ^3 } \right)}$ where $\gamma=26.7522$ ($10^{7}$ rad/sT) is the gyromagnetic ratio for $^{1}$H, and $D$ the water self-diffusion coefficient ($D=2.3\times10^{-9}$ m$^{2}$/s for bulk water at room temperature). The linearly exponential part of the decay corresponds to the $^1$H NMR spin-spin T$_2$ relaxation, while the cubic exponential decay corresponds to $^1$H NMR spin-echo dephasing in the magnetic field gradient G, in the presence of diffusional motion. In the case of non-uniform diffusion processes with multiple $D$ values, diffusion is expressed with a distribution function $f(D)$, which is obtained by implementing an appropriate  2D inversion algorithm \cite{mitchell2012numerical}, as explained in the Supporting Information. It is noticed that $^{1}$H NMR studies of water diffusion inside CNTs have been already reported, presenting mainly effective $D$ values \cite{das2010single,liu2014diffusion,Hassan2016Review}. However, it is extremely difficult with conventional 1D NMR diffusion experiments to disentangle water components with overlapping $D$ coefficients. To overcome this obstacle, we take benefit of the fact that nanotube water, interstitial water, and bulk water, acquire different $T_2$ values, as already demonstrated in Figure~\ref{fig:fig1}b. Diffusion differences can be thus acquired by analyzing  signals in different $T_2$ windows. This intricate assignment of diffusion differences in T$_2$ resolved water groups was excellently demonstrated in the past, through $^1$H NMR study of water mobility in hydrated collagen II \cite{knauss1996pulsed}. \ 

Fig. \ref{fig:fig2} shows the 2D \(D-T_2\) contour plots of water in SWCNT and DWCNT  at selected temperatures. Signals at different temperatures have been scaled accordingly to improve visualization. Two main peaks are observed, representing water inside the CNT channels (nanotube water) with very short $T_2$  at $\approx0.5$ ms and external water with long $T_2$ at $\approx10-12$ ms. A detailed presentation of the dependence of $T_2$ on the temperature is presented in the Supporting Information. It is noticed that short $T_2$ values might create distortions in the $D-T_2$ spectra, therefore the validity of the inversion algorithm has been confirmed by comparing experimental with simulated spectra, as shown in the Supporting Information.

In both CNTs, external water attains $D$ value $\approx2.5\times10^{-9}$ m$^{2}$/s at room temperature, which is close to that of bulk water. Upon lowering temperature, water mobility decreases and the signal shifts to lower $D$ values, as expected. At the same time, the intensity of the NMR  signal of the external water decreases rapidly. We notice that part of the signal persists even at temperatures lower than the typical freezing point of bulk water ($\approx273$K), in agreement with previous $^1$H MAS NMR measurements \cite{Sekhaneh2006}. Therefore, the signal of the long $T_2$ water component, which remains at temperatures below the water freezing temperature, cannot be assigned to ``bulk'' water but rather to interstitial water confined in the space between the CNT bundles (confined water is known to freeze at temperatures lower than the freezing temperature of bulk water \cite{ashworth1984}).\\
\indent
The $D$ projections shown at the left side of the contour plots in Fig. \ref{fig:fig2} have a contribution from both interstitial and nanotube water signals; the diffusion peak from water inside the CNTs is distorted or even masked by the signal from the external water. 
\begin{figure}[ht]
\centering
\includegraphics[width=\linewidth]{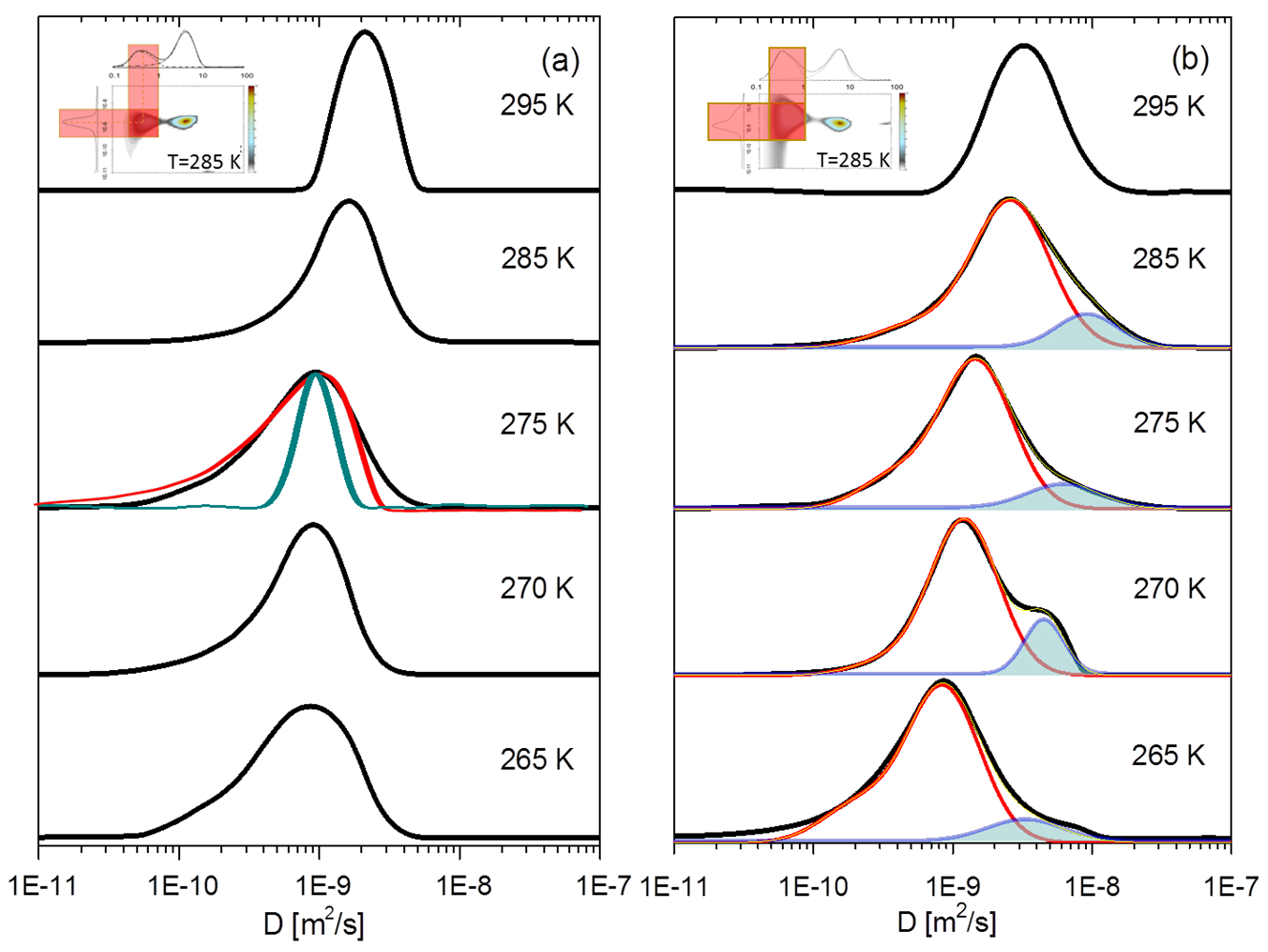}
\caption{ $D$-projections of nanotube water in (a) SWCNTs and (b) DWCNTs. The insets show the selected areas of projection. For better visualization spectra are normalized to one. The red line at 275 K in SWCNT is simulation curve of water diffusion in a bundle of randomly oriented CNTs (see text). The dark cyan line is the relevant simulation curve of unrestricted bulk water diffusion. In DWCNTs nanotube water shows stratified fast diffusion. The red line corresponds to shell water as in SWCNTs, while the cyan  line represents the axial fast nanotube water component.}
\label{fig:fig3} 
\end{figure}
Hence, in order to uncover the water dynamics inside the CNT channels,  the $D$ projections of nanotube water from both SWCNTs and DWCNTs were separately calculated, as presented in  Fig. \ref{fig:fig3}. 

In the case of SWCNTs ( left panel of the figure), at temperatures below $285$ K, the $D$ projection exhibits the characteristic low-$D$ tail, encountered with 1D diffusion in randomly oriented CNTs \cite{callaghan1979diffusion}. This is manifested with the theoretical red line  (details are provided in the Supporting Information), which simulates accurately the experimental spectrum at 275 K. Fitting of the curve to the 1D model with a unique $D$ value is consistent with the confinement of water molecules in a randomly oriented array of CNT tubes with short (nanometer scale) inner radial dimension but long (micrometer scale) axial dimension.  Most important, the restricted diffusion in SWCNTs acquires $D$ values close to that of free water, in agreement with MDS \cite{alexiadis2008molecular}.
\begin{figure*}[!ht]
\centering
\includegraphics[width=\linewidth]{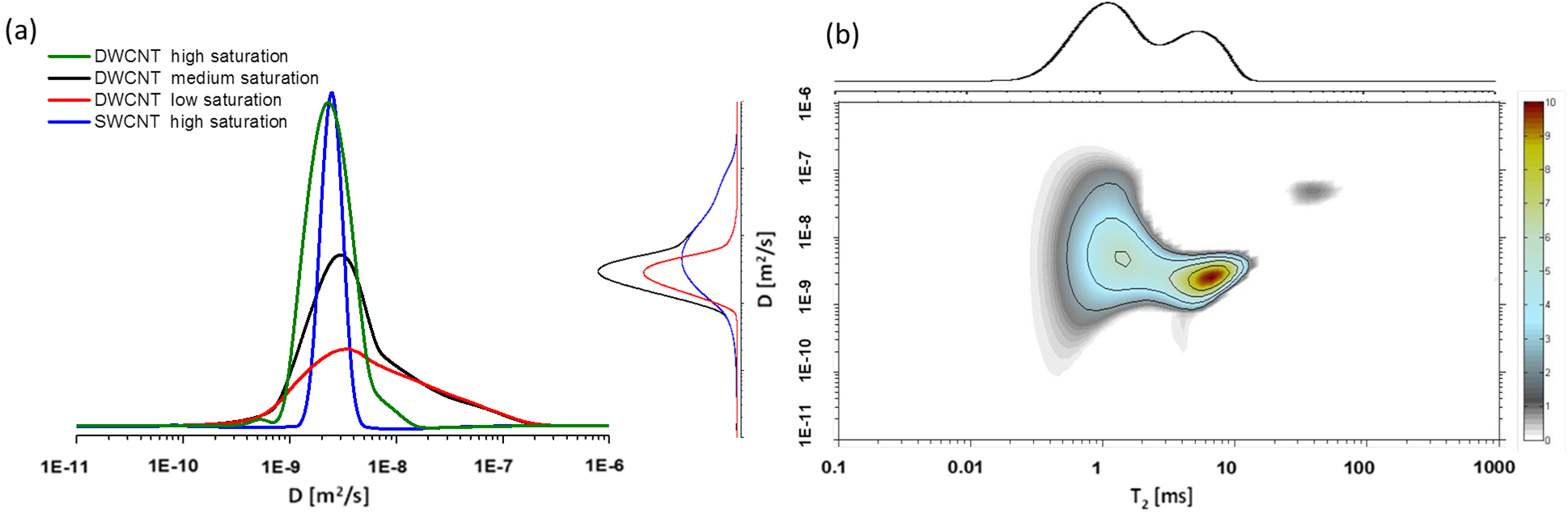}
\caption{ (a) The $D$-profile of water in SWCNT and DWCNTs, at room temperature at different water content. For better visualization the spectra are re-scaled accordingly. Upon lowering water content, the profile moves toward higher values of $D$, in agreement with MD simulation results  \cite{striolo2006mechanism}. (b) The $^{1}$H NMR $D-T_2$ spectrum of  DWCNTs with intermediate water content  (corresponding to the black line in pannel (a). }
\label{fig:fig4}
\end{figure*}
In DWCNTs (right panel of Fig. \ref{fig:fig3}), the $D$ distribution of water inside the CNTs attains a more intriguing picture. At $T=285$ K, the diffusion profile of the nanotube water can be resolved into two components having different dynamics with $D$ peaks centered at $\approx2.58\times10^{-9}$ m$^{2}$/s, and $9.60\times10^{-9}$ m$^{2}$/s.  This is in agreement with neutron scattering experiments, which confirm the presence of two different water groups, i.e.  a shell of water close to the CNT walls and a chain-water at the center of the CNTs \cite{kolesnikov2004anomalously}. Furthermore, MDS have shown a stratified arrangement of water molecules inside CNTs \cite{bordin2013relation,joseph2008carbon,alexiadis2008molecular,barati2011spatial}. The stratified arrangement of water in the DWCNTs might be explained by considering (i) the larger inner size of $3.5$ nm compared to that of SWCNT, and (ii) the repulsive Coulomb forces \cite{head1993orientational} between next neighboring oxygen atoms, which become important when water is squeezed into  hydrophobic CNTs. Evidently, water inside DWCNTs is attributed to shell and chain-water, the latter shows astonishing diffusion enhancement, which becomes $\approx8-9$ times higher than that of bulk water at room temperature. This behaviour was not observed in SWCNT due to the small confinement size of $\approx1 nm$. 

More to this discussion, we noticed the diffusion of water molecules is further enchanced upon decreasing water content in the CNT samples, in agreement with the theoretical predictions \cite{striolo2006mechanism}. Fig. \ref{fig:fig4} shows the $D$-distribution profile of water, at room temperature, in SWCNT as well as in three DWCNTs samples with different water content. Results show that by reducing water, a very broad distribution of $D$ unveils, which decomposes into a very broad fast diffusing component belonging to the nanotube water and a second one belonging to the external water. 
\begin{figure}[!ht]
\centering
\includegraphics[width=\linewidth]{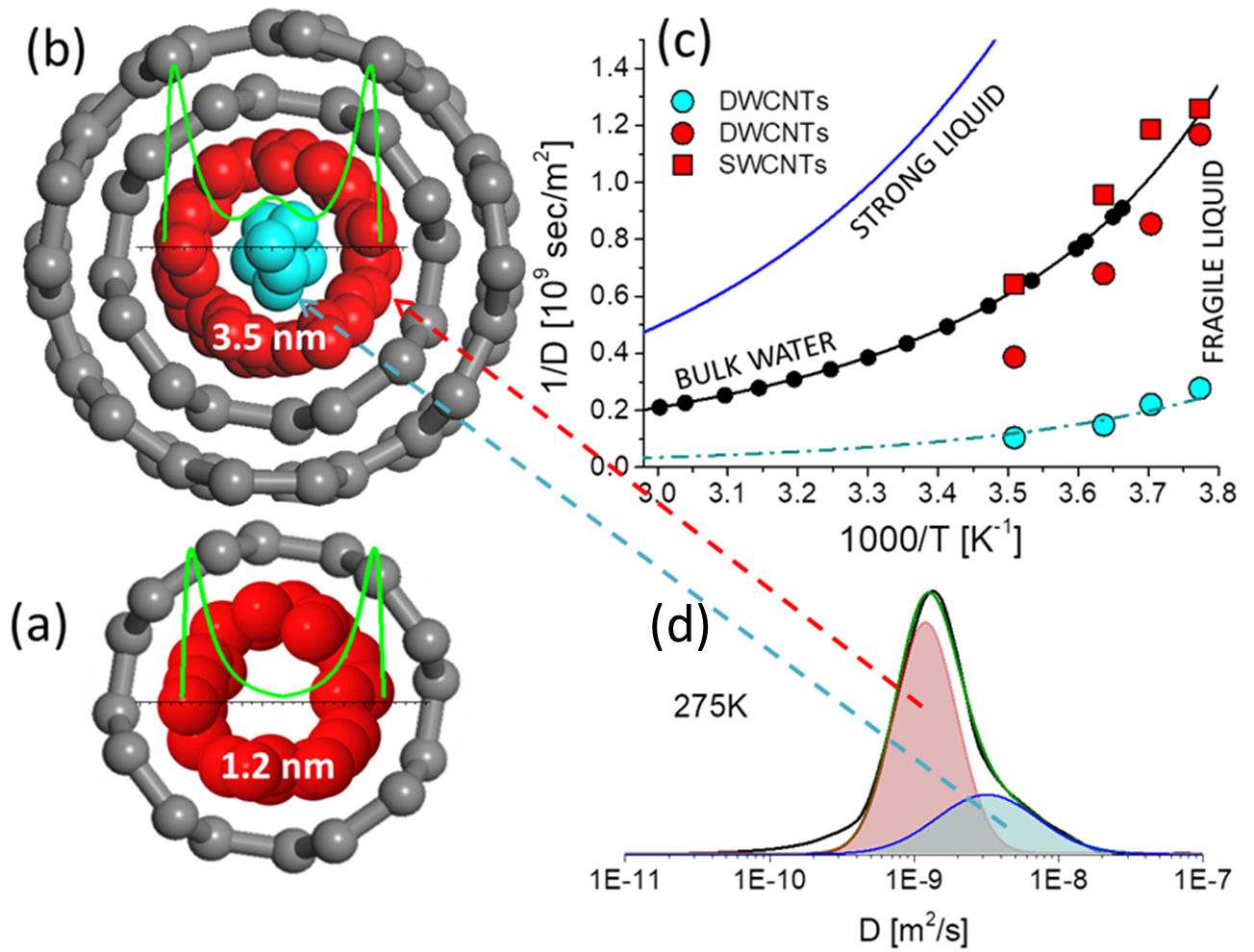}
\caption{Schematic presentation of water diffusion in SWCNTs (a) and DWCNTs (b). Water structure has been sketched by combining \(D-T_2\) NMR experimental data with recent MDS \cite{bordin2013relation,joseph2008carbon}. The green peaks in pannels (a) and (b) are sketches of the relevant water density.  (c) The inverse of the experimental self-diffusion coefficient $1/D$ as a function of $1000/T$. The blue line shows the temperature dependence of an ideal ``strong'' liquid following Arrhenius law. The solid black and the dashed cyan lines are theoretical fits as described in the text. (d) $^1$H NMR $D$ distribution of water in a DWCNTs, at $275$K. The cyan and red color shaded areas, correspond to the axial inner (cyan color) and outer (red color) water components of pannel (b). }
\label{fig:fig5}
\end{figure}

To summarize the results, a schematic representation of the water groups in SWCNTs and DWCNTs is shown in Fig. \ref{fig:fig5}. According to the NMR experiments, in SWCNT with inner diameter $\approx1.2$ nm,  only a single shell nanotube water component was detected (represented by the red spheres in Fig. \ref{fig:fig5}a), in agreement with the MDS \cite{bordin2013relation,joseph2008carbon}. Water molecules of this group have self-diffusion coefficients (red squares in Fig. \ref{fig:fig5}c) comparable to that of bulk water. Remarkably, in DWCNTs with inner diameter $ \approx3.5$ nm, in addition to the shell water, a second water component is observed (cyan spheres in Fig. \ref{fig:fig5}b), with sufficiently higher $D$ values in comparison to those of bulk water, as seen in Fig. \ref{fig:fig5}c. The green lines in Figures \ref{fig:fig5}a,b represent water profiles according to MDS \cite{bordin2013relation,joseph2008carbon}.  These findings confirm theoretical calculations, which depending on the CNT-size, predict that water inside CNTs diffuses in the shape of hollowed nanotubular clusters with an ultrafast mechanism \cite{striolo2005water,striolo2006mechanism}. The lower $D$ values of the outermost axial water component in both SWCNT and DWCNT may be attributed to ``friction'' from the interaction of water molecules with defects and oxygen moieties, which are unavoidably present on the CNT walls. 

In order to further enlighten the dynamics of the nanotube water in both CNTs, the inverse of the self-diffusion coefficient, $1/D$ versus $1000/T$  is presented in Fig. \ref{fig:fig5}c. For comparison, $1/D$ of bulk water (black circles) is shown in the same Figure. The blue line is the theoretical simulation of an ideal ``strong'' liquid with $1/D$ exhibiting Arrhenius temperature dependence, $\frac{1} {{D}}\propto \:exp{\left ({\frac{U}{k_BT}} \right)}$. Bulk water is observed to deviate significantly from the Arrhenius behavior; Within Angels concept of fragility \cite{speedy1976isothermal}, bulk water is considered to be fragile \cite{prielmeier1987diffusion,dehaoui2015viscosity}, i.e. it slows down gradually upon cooling and freezes rapidly  near the water glass transition temperature at $T_g \approx 40$K. The black line is the theoretical fit to the empirical formula  $\frac{1}{{D}}\propto \frac{1}{{D_0}}{(\frac{T-T_S}{T_S})}^\gamma$, which is widely used in the literature to fit the non-Arrhenius temperature dependence of the water transport properties \cite{speedy1976isothermal}. $T_S$ in this formula is the thermodynamic limit where transport properties become zero, and the exponent $\gamma$ provides a measure of the growth of the hydrogen bonding by cooling. Experimental data are nicely fitted with \(\gamma\approx2\) and \(T_S\approx215\) K, in agreement with previous reported values \cite{perakis2010two}. In the case of SWCNT, the $1/D$ values of nanotube water matches  well with those of bulk water. However, the $1/D$ values of the central water component  in the DWCNTs  (cyan circles in Fig. \ref{fig:fig5}c) deviate strongly from those of bulk water, acquiring low values which vary slowly by decreasing temperature. This indicates that the central nanotubular water component is more fragile, and "resists" in the formation of hydrogen bonds upon cooling \cite{bordin2013relation,mauro2014structural}.  It is possible that at room temperature, liquid islands in this fragile water component are responsible for the weak "liquid-like" $T_1$ component at $\approx1.2$ s, which is observed in the left inset of  Fig. \ref{fig:fig1}a (cyan line). 
\\
These findings provide a different viewpoint to the way that water is organized inside the CNTs.   It is generally accepted that water mobility in CNTs depends strongly on the number of hydrogen bonds per water molecule and the inner diameter of the CNTs \cite{barati2011spatial,joseph2008carbon}. Specifically, MDS have shown that the number of hydrogen bonds per water molecule is $\approx3.7$ at the center of the nanotubes (the same as bulk water), and $\approx2$ close to the CNT walls, which gives rise to enhanced water mobility close to the hydrophobic CNT walls \cite{barati2011spatial,joseph2008carbon,Gogotsi2001}. This is in contrast to the results presented here, which show that water diffusion at the center of DWCNTs is sufficiently faster than diffusion near the walls. The observed reduced mobility close to the walls might be related to the ``friction'' caused by the interaction of the water molecules with defects and oxygen moieties, referred above. Furthermore, our experimental results are in agreement with viscosity measurements \cite{kohler2016}, which show that shear viscosity of water confined in CNTs is about an order of magnitude lower than in bulk water and increases non-linearly with CNT diameter, while enhanced diffusion of water inside CNTs appears only for diameters higher than $4$ nm. 
 
\section*{Conclusions}
\indent
In conclusion, herein we report the first experimental measurement of fast water diffusion in single and double wall CNTs with the help of 2D $^{1}$H NMR (\(D-T_2\)) and 2D $^{1}$H NMR (\(T_1-T_2\)) methods. Unlike previous works where diffusion of water in CNTs at the molecular level has been monitored almost solely with MDS, here we demonstrate a fast and scalable method to acquire experimentally diffusion of water and other fluids in nanotubular channels at nanoscale resolution. With this method, water in DWCNTs is shown to have a stratified diffusion profile, revealing an axial component with extraordinary high $D$ values, a phenomenon not present in SWCNTs. The discovery of stratified fast diffusion in CNTs is new experimental evidence, which confirms MDS predictions, with important impacts on our current understanding of water flow through hydrophobic nanochannels and nanostructured membranes \cite{garcia2012designing,geng2014stochastic}. It is also expected to be very helpful in the efforts to understand fluidic properties  through extremely narrow pores, with the aim to develop high readout single molecule detectors \cite{liu2010translocation,liu2013ultrashort}, and regulate the cellular traffic of important biological solutes \cite{fornasiero2008ion}. 

\section*{Methods}

\indent
Purified and open ended single and double walled carbon nanotubes, SWCNT and DWCNT, were purchased from SES research, USA. The inner diameter of the CNTs used in this work were $\approx1.2$ nm for the SWCNT and $\approx3.5$ nm for the DWCNT. Both types of CNTs had average length $\approx1$ $\mu$m. \ 
The samples were characterized using a TEM-FEI Tecnai G20 instrument with a $0.11$ nm point to point resolution. More information on the samples and their TEM images are available in the Supporting Information.\\
For the NMR experiments, CNT powders were used with no further treatment and double distilled water was used in all the measurements. $^1$H NMR two-dimensional diffusion-relaxation 2D $D-T_2$ measurements were performed in the stray field of a $4.7 $ T Bruker superconductive magnet providing a $34.7$ T/m constant magnetic field gradient at $^1$H NMR frequency of $101.324 $MHz. The experiments were carried out by using a pulse sequence with more than $5000$ pulses (more detail can be found in Supporting Information). The temperature was controlled by an ITC5 temperature controller in a flow type Oxford cryostat. The accuracy of the temperature was $\approx0.1$ K. A thirty-minute time window was allowed at each temperature before collecting data. 
NMR data were analyzed using a 2D non-negative Tikhonov regularization inversion (discussed in the Supporting Information) algorithm code, developed by the authors.
The $^1$H Magic Angle Spinning (MAS) NMR experiments were performed at room temperature on a BRUKER (AVANCE 400) NMR spectrometer at spinning frequency of $12$kHz.

\acknow{This work was supported by Khalifa University Fund (210065). G.P. and 
H.J.K. would like also to express their gratitude to IRSES FP7 project Nanomag (295190)}
\showacknow{} 

\bibliography{pnas-refcnt}
\bibliographystyle{pnas-refcnt}
\end{document}